\documentclass[conference]{IEEEtran}

\usepackage{times}
\usepackage{epsfig,graphicx,color,pstricks}
\usepackage{amsmath,amssymb,amsthm,amsfonts,eucal,latexsym}

\usepackage{cite}       

\usepackage{graphicx}   

\usepackage{psfrag}    

\usepackage{subfigure} 

\usepackage{url}       

\usepackage{stfloats}  

\usepackage{amsmath}   
\usepackage{array}
 \makeatletter
 \let\NAT@parse\undefined
 \makeatother
\usepackage{epsfig,graphicx,color,pstricks}
\usepackage{amsmath,amssymb,amsthm,amsfonts,eucal,latexsym,dsfont}
\usepackage{cite}
\usepackage{amsmath,amssymb,eucal,graphicx}
\usepackage{pst-plot}
\usepackage{amsthm}
\usepackage{amsfonts}
\usepackage{psfig}
\usepackage{epsfig}
\usepackage{float}
\usepackage{pstricks}
\usepackage{color}

\begin{document}

\title{On Repetition Protocols and Power Control for Multiple Access Block-Fading Channels}

\author{
\IEEEauthorblockN{Davide Barbieri and Daniela Tuninetti,
}
\IEEEauthorblockA{University of Illinois at Chicago (UIC),
dbarbi2@uic.edu and danielat@uic.edu}
}



\maketitle

\begin{abstract}
In this paper we study the long-term throughput performance of repetition protocols coupled with power control for multiple access block-fading channels. We propose to use the feedback bits to inform the transmitter about the decoding status {\em and} the instantaneous channel quality. We determine the throughput of simple and practically inspired protocols; we show remarkable throughput improvements, especially at low and moderate SNR, when compared to protocols where the feedback bits are used for acknowledgment only or for power control only; we show that the throughput is very close to the ultimate ergodic multi-user water-filling capacity for small number of feedback bits and/or retransmissions.  For symmetric Rayleigh fading channels, numerical results show that the throughput improvement is mainly due to the ability to perform a power control, rather than to retransmit.
\end{abstract}

\section{Introduction}
\label{sect:intro}

In current networks, reliability 
is obtained with a combination of FEC (Forward Error Correction) 
and HARQ (Hybrid Automatic Repetition reQuest). 
FEC attempts to correct transmission errors by using
error correcting codes (convolutional codes, turbo codes, etc.),
while HARQ protocols request a retransmission 
when an error
is detected.  HARQ protocols require a feedback channel to indicate a
decoding success/failure 
to the transmitter. In wireless channels
without CSI (Channel State Information), HARQ provides time diversity and robustness
to channel variations~\cite{book:goldsmith:wireless}.  If CSI is available
at the transmitter, adaptive rate and power allocation becomes possible~\cite{book:goldsmith:wireless}. 
Although power control improves the performance of single-user channels
at low SNR (Signal to Noise Ratio) only, it is well known to provide
multi-user diversity in MAC's (Multiple Access Channel)~\cite{knopp_humblet:multiuser_waterfilling}.
The central question of this paper is: with the objective to
maximize the network throughput, should the available feedback resources
be used for HARQ, or to provide CSI?  
In this paper, we consider fading MAC's.
We propose that the available feedback bits are used to communicate
to the transmitter {\em both} the decoder state and the channel state,
as proposed in~\cite{tuninettiICC2008} for single-user channels.  
Our goal is to simultaneously realize the gains of {\em power control}
and of {\em repetition protocols} by using the
same feedback resources that would be used for repetition alone.

We assume that the CSI (i.e., the fading gains from the transmitters to the receiver)
is perfectly known at the receiver, for example through pilot tones.
The receiver then informs
the transmitters about the channel {\em and} the decoder state through a 
common (broadcast) feedback channel of finite capacity.  Since the feedback channel
is rate-limited, the transmitters only have partial CSI, based on which they need
to decide an appropriate power level for the (re)transmission.  
Past work 
considering limited CSI
only focused on ergodic capacity or on outage probability, but not
on HARQ protocols--which is the goal of this paper.
For example, the work in~\cite{Bhashyam} showed that power control is very useful
to minimize the outage probability even with partial and/or noisy CSI,
while the work in~\cite{tse_hanly_p1} derived the ergodic capacity
of the MAC with perfect CSI.  


Here we extend the joint HARQ and power control protocol
proposed in~\cite{tuninettiICC2008} for a single-user channel to MACs.
The protocol in~\cite{tuninettiICC2008}
is equivalent to a time-varying (depending on the current retransmission attempt)
quantizer of the amount of redundancy that is still needed at the
decoder for successful decoding.
The novel technique of~\cite{tuninettiICC2008}
improves the long-term average throughput at any SNR,
compared to schemes with power control and/or HARQ alone;
moreover, it achieves at least 67\% of the ergodic water-filling capacity
with a single retransmission and one~bit of feedback.
The ergodic water-filling capacity is the fundamental performance limit
of a fading channel with perfect CSI and with the possibility to code
across many channel coherence times, i.e., it is achievable
when a feedback channel of infinite capacity is available 
(to provide perfect CSI to the transmitter)
and there is no upper limit to the number of retransmissions
(to ensure zero outage).

In this paper, we consider a $K$-user block-fading Gaussian MAC
where each user has $M$ transmission attempts to send a data packet to
the central receiver.  The receiver is assumed to have an error-free 
broadcast feedback channel of capacity of $\log_2(F)$~bits to communicate to
the transmitters.  Our performance measure is
the {\em long-term achievable throughput} (or simply throughput for short in the following)
subject to a {\em long-term power constraint}
at each transmitter~\cite{caire_taricco_biglieri:optimal_powercontrol,
tuninettijouiranlACKvsCSI}.  This performance measure includes the
outage capacity~\cite{summa_fading_bc} and the ergodic capacity~\cite{knopp_humblet:multiuser_waterfilling}
as a special case for $M=1$ and $M=\infty$, respectively~\cite{caire_tuninetti:arq_it}.
For a finite ($M,F$), deriving the optimal throughput,
or an outer bound, seems very difficult. For this reason, we
propose some simple and practically inspired
schemes, we evaluate their throughput, and compare them
with the ultimate ergodic water-filling capacity (i.e., case $(M,F)=(+\infty,+\infty)$).

The rest of the paper is organized as follows: Section~\ref{sect:model}
introduces the system model; Section~\ref{sect:M=1} proposes various policies
for $M=1$; Section~\ref{sect:M=2} discusses the case of $M\geq2$;
Section~\ref{sect:num} reports numerical 
results, and Section~\ref{sect:concl} concludes the paper.

\section{System Model}
\label{sect:model}

{\bf Notation:}
$\mathbb{E}[X]$ is the expected value and
$F_X(x)$ 
is the cumulative distribution function (cdf) of the random variable $X$.
$\mathbb{P}[A]$ is the probability of the event $A$.
$\mathds{N},\mathds{R}$ and $\mathds{C}$ indicate, respectively, the natural, the real and the complex numbers.

Plots show the ratio of throughput (finite ($M,F$))
and ergodic capacity ($(M,F)=(+\infty,+\infty)$) vs SNR in dB.
The numerical results are for unit power Rayleigh fading,
i.e., the fading power has cdf $F_X(x)=1-e^{-\max\{x,0\}^+}$, 
for which the 
$\mathbb{E}[1/X \cdot 1_{\{X\geq x  \}} ]
={\rm Ei}(x)=\int_x^\infty e^{-t}/t \ {\rm d}{t}$, $x\geq 0$.
Equalities labeled with $\star$ hold for
for a {\em symmetric} system
where all users have the same power constraint, the same transmit rate, and
the same fading iid Rayleigh fading; in this case we shall drop the index of
the user.

\smallskip
{\bf Modeling Assumptions:}
We consider  a block-fading Gaussian MAC. 
The receiver knows perfectly the channel fading at the beginning of each slot, whereas the transmitters have no CSI,
unless explicitly informed by the receiver.  Transmitters can not modify the rate of communication in each slot
and cannot send a superposition of different codebooks, as opposed to~\cite{Kim-Skolunt} and to~\cite{steinber-samai}
which considered multi-layer transmission.  Power allocation is instead permitted, exploiting the partial CSI obtained from the receiver.
Each transmitted codeword spans one fading block, or slot, over which the fading is constant. The slot length is sufficient to permit successful decoding if the mutual information at the receiver is above the transmission rate~\cite{summa_fading_bc}. 
%
The broadcast feedback channel is assumed error-free, delay-free and of finite capacity given by $\log_2(F)$, $F\in\mathds{N}$.
Accumulation of feedback bits over successive slots is not permitted.
The receiver can detect uncorrectable errors and in this case can ask for a retransmission.
Each transmitter can transmit at most $M$ times, $M\in\mathds{N}$, the same data packet, including the first transmission.
The performance measure is the \textit{long-term average throughput} vs. the \textit{long-term average power}.
Long-term means that that the averages are evaluated over a time horizon much larger than $M$.
Power control permits to use more power when the instantaneous channel conditions are ``good'', as long as
the average power constraint is not violated.  Peak power constraints, although important in practice, are not considered in this paper and are left for future work.

\smallskip
{\bf Performance Measure:}
Under these assumptions, the received signal in slot $t \in \mathds{N}$ of a $K$-user MAC is:
\begin{equation}
\textbf{Y}_t = \sum_{k=1}^{K} h_{k,t}\sqrt{P_{k,t}} \textbf{X}_{k,t} + \textbf{Z}_t \in \mathds{C}^{n\times 1}
\end{equation}
where: 
$h_{k,t}$ is the fading gain for user $k$,
$\textbf{Z}_t$ is the AWGN with zero mean and unit variance,
$\textbf{X}_{k,t}$ is the Gaussian codeword 
of length $n$, such that $1/n\ \mathbb{E}[||\textbf{X}_{k,t}||^2 ] = 1$,
and $P_{k,t}$ is the instantaneous power that must satisfy the power constraint $\overline{P}_{k}$,
that is, $\lim_{T\to\infty}1/T\sum_{t=1}^TP_{k,t} \leq \overline{P}_{k}$.
We assume $P_{k,t}\in\{P^{(1)}_k, ..., P^{(M)}_k\}$, where $P^{(m)}_k$ is the power policy to be used at the $m$-th transmission attempt, $m=1,..,M$, that can take at most $F$ values.

Let $R_{k,t}$ be the rate decoded at time $t$ for user $k$. The throughput is 
$\eta_{M,F,K}=\lim_{T\to\infty}1/T\sum_{t=1}^T (R_{1,t}+...,R_{K,t})$.
For any $(M,F)$, the throughput is upper bounded by the ergodic water-filling capacity (ewfc) 
\begin{align}
\eta_{M,F,K}
  &\leq \eta_{K}^{\rm(ewfc)}  
\stackrel{\star}{=} \sum_{k=1}^K (-1)^{k-1} {K \choose k} {\rm Ei}(k x), 
\label{eq:erg}
\end{align}
where $x\geq 0$ in~\eqref{eq:erg} is linked to the power constraint  through
\begin{align}
\overline{P} \stackrel{\star}{=}
\sum_{k=1}^K (-1)^{k-1} {K \choose k}\Big(e^{-k x}- k x\ {\rm Ei}(k x)\Big).
\end{align}

\section{Case $M=1$: Outage Capacity}
\label{sect:M=1}

In the following we introduce protocols/policies
for the case where only one transmission per data packet is allowed (i.e. $M=1$).
Policies are divided into two categories: a) those that do not exploit CSI
and, therefore, are not able to perform power adaptation, 
and b) TDMA-type policies, where the user with the largest fading gain
is allowed to transmit at any give time (thus requiring CSI).
Although the proposed policies are not throughput optimal, we will show that they
achieve a large fraction of the ergodic water-filling capacity $\eta_{K}^{\rm(ewfc)}$ in~\eqref{eq:erg}.

\smallskip
{\bf Policies without CSI (i.e. $F=1$):}
Let parametrize the rates as $R_u = \log(1+s_u P_u)$ for some $s_u\geq 0$, $u=1,...,K$,
to be optimized.
Consider a \underline{static TDMA scheme}, where each user sends for a fraction $1/K$ of the time
with power $P_u = K \overline{P}_u$. The throughput is
\begin{align}
&\eta_{M=1,F=1,K}^{\rm (static-TDMA)}
=\max_{s_1,...,s_K} \sum_{k=1}^{K}\frac{1}{K}\log\left(1+s_k\,K \overline{P}_k\right)
      \mathbb{P}[|h_k|^2>s_k] \nonumber
\\&\stackrel{\star}{=} \max_{s\geq0} \left\{e^{-s}\log\left(1+s\,K \overline{P}\right)\right\}
=\eta_{\rm single}(K \overline{P}).
\label{eq:tdma_fixed}
\end{align}

\smallskip
If the transmitters are allowed to transmit simultaneously, then the receiver must perform
\underline{joint decoding.} For the case of $K=2$ users
(extensions to more than two users is straightforward) we get
\begin{align}
&\eta_{M=1,F=1,K=2}^{\rm(joint)}
   = \max_{R_1,R_2}\{R_1\cdot \mathbb{P}_{10} + R_2\cdot \mathbb{P}_{01} +(R_1+R_2)\cdot \mathbb{P}_{11} \}\nonumber
\\&\stackrel{\star}{=} \max_{s\geq 0} \Big\{2 \  e^{-s}\
\left(\frac{1-e^{-s(\overline{P} s+1)}}{\overline{P} s+1}
     +e^{-s(\overline{P} s+1)}(\overline{P} s^2+1)\right)
 \cdot \nonumber
\\&\quad\cdot
     \log(1+\overline{P} s) \Big\}=\eta_{\rm jont}(\overline{P}),
\label{eq:joint_nocsi}
\end{align}
where
\begin{equation*}
\mathbb{P}_{11}=\mathbb{P}\left[\begin{array}{cc}
       R_1 \leq \log\left(1+|h_{1,t}|^2P_{1,t}\right) \\
       R_2 \leq \log\left(1+|h_{2,t}|^2P_{2,t}\right) \\
       R_1+R_2 \leq \log\left(1+|h_{1,t}|^2P_{1,t}+|h_2|^2P_{2,t}  \right)
       \end{array}\right] 
\end{equation*}
is the probability that the two users can be jointly decoded,
\begin{equation*}
\mathbb{P}_{10}=\mathbb{P}\left[\begin{array}{cc}
      R_1 \leq \log\left(1+\frac{|h_{1,t}|^2P_{1,t}}{1+|h_{2,t}|^2P_{2,t}}\right) \\
      R_2 > \log\left(1+|h_{2,t}|^2P_{2,t}\right)
      \end{array}\right] 
\end{equation*}
is the probability that user~1 can be decoded by treating user~2 as noise and that 
user~2 cannot be decoded, and $\mathbb{P}_{01}$
is as $\mathbb{P}_{10}$ but with the role of the users swapped. In the case 
$M=F=1$, we have $P_{u,t} = \overline{P}_u$, $\forall (t,u)$ (because of neither power
adaptation is possible, nor repetition).

\smallskip
Finally, we consider a hybrid version of the two previous policies with
\underline{TDMA and joint decoding}.
For the symmetric case (the extension to the non-symetric case is straightforward):
divide the communication frame into three parts.
Each user transmits alone for a fraction $\frac{\tau}{2}$ of the time with 
power $\frac{2\alpha}{\tau}\overline{P}$, $\tau\in[0,1],\alpha\in[0,1]$;
both users transmit together for a fraction  $1-\tau$ of the time
with power $\frac{1-\alpha}{1-\tau}\overline{P}$.
The throughput is
\begin{align}
    \eta_{M=1,F=1,K=2}^{\rm(joint+TDMA)}
&\stackrel{\star}{=}
   \max_{\tau\in[0,1],\alpha\in[0,1]}
   \Big\{\tau\ \eta_{\rm single}\left(\frac{2\alpha}{\tau}\overline{P}\right)+ \nonumber\\
    & +(1-\tau)\ \eta_{\rm joint} \left(\frac{1-\alpha}{1-\tau}\overline{P}\right)\Big\},
\label{eq:joint_single}
\end{align}
with $\eta_{\rm joint}$ in~\eqref{eq:joint_nocsi} and $\eta_{\rm single}$ in~\eqref{eq:tdma_fixed}.

\smallskip
{\bf TDMA-type policies with partial CSI (i.e. $F>1$):}
The following schemes are inspired by the ``channel-driven TDMA'' (cdTDMA)
power allocation that achieves the ergodic water-filling capacity in~\eqref{eq:erg}.

In the first scheme, the user with the largest fading gain is allowed to
transmit at any given time with the maximum possible power (\underline{cdTDMA-on}).
The feedback bit indicates the user who is allowed to transmit, that is,
\[
B = \arg\max_{k=1,...,K}\{|h_k|^2/s_k\},
\]
for some $s_k\geq 0$, $k=1,...,K$, to be optimized.
This scheme requires $F=K$ values of feedback (i.e., the number of feedback bits per user
is $1/K \log_2(K) \to 0$ as $K\to\infty$).
The rate of user $k$, $k=1,...,K$, is parameterized as $R_k=\log(1+s_k P_k)$,
where $P_k=\overline{P}_k / \mathbb{P}[B=k]$ is the transmit power
when the user is active. The throughput is 
\begin{align}
  & \eta_{M=1,F=K,K}^{\rm (cdTDMA-on)}\nonumber
\\&= \max \sum_{k=1}^{K}\log\left(1+\frac{s_k\,\overline{P}_k}{\mathbb{P}[B=k]}\right) 
  \mathbb{P}[s_k < |h_k|^2, B=k] \nonumber
\\&\stackrel{\star}{=}
 \max_{s\geq 0}  \log\left(1+s\,K\overline{P}\right) [1-(1-e^{-s})^K].
\label{eq:tdma_best}
\end{align}

\smallskip
The second policy is a modification of the previous one. The difference is that the user with
the largest fading gain is allowed to transmit only if the fading is above a certain threshold,
i.e., power can be switched off in deep fade (\underline{cdTDMA-on/off}).
Let the feedback be
\begin{equation*}
B = u \ {\rm if} \ 
\left\{
u = \arg\max_{k=1,...,K}\frac{|h_k|^2}{\lambda_k} \
{\rm and} \ |h_u|^2>\tau_u
 \right\},
\end{equation*}
and $B=0$ otherwise.
User~$u$ sends in the event $B=u$ and uses power
$P_u=\frac{\overline{P}_u}{\ \mathbb{P}[B=u]}$, otherwise it is silent.
This policy requires $F=K+1$ values of feedback.
The throughput is
\begin{align*}
&\eta_{M=1,F=K+1,K}^{\rm (cdTDMA-on/off)}
  = \max_{\{R_u,\lambda_u,\tau_u\}}
  \sum_{u=1}^{K} R_u \nonumber\\
 & \cdot \mathbb{P}\left[R_u <\log\left(1+|h_u|^2\frac{\overline{P}_u}{\ \mathbb{P}[B=u]}\right), B=u\right]\nonumber
\end{align*}
In the symmetric case it can be shown that  $\lambda_u=\tau_u=s$ for all $u=1,...,K$ is optimal and that
\begin{align}
&\eta_{M=1,F=K+1,K}^{\rm (cdTDMA-on/off)}\nonumber
\\&\stackrel{\star}{=}  
\max_{s\geq0}      \log\left(1+\frac{sK\overline{P}}{1-(1-e^{-s})^K}\right)[1-(1-e^{-s})^K],
\label{eq:TDMA_thr}
\end{align}
It is immediate to see $\eta_{M=1,F=K,K}^{\rm (cdTDMA-on)}\leq\eta_{M=1,F=K,K}^{\rm (cdTDMA-on/off)}$.

\smallskip
We could also consider ``hybrid'' policies where joint decoding and cdTDMA coexist
(\underline{joint+cdTDMA}). 
In~\cite{davidemsthesis}, we analyzed such more complex power policies for $M=1$; it turned out that the improvement over cdTDMA-on/off is minimal and does not justify the increased in complexity.

\smallskip
The policies discussed so far used are characterized by $\log_2(F) \leq K$, that is, the amount of feedback resources per user is the same as that of a simple HARQ protocol.  We next show how to get closer to the ergodic water-filling capacity by allowing $\log_2(F) > K$.  We shall see that the ability to perform a finer channel adaptation (due to a better CSI) greatly increase the achievable throughput. The previous cdTDMA-on/off policy allowed for the instantaneous power to be either zero or $P_1>0$ and required $F=K+1$ feedback values. We extend it now so that the instantaneous power can be from the set $\{0,P_1,...,P_L\}$ with $0<P_1<...<P_L$ thus requiring $F=LK+1$ feedback values.  We refer to this policy as \underline{multilevel cdTDMA-on/off}. We only discuss the symmetric systems; the extension to a general setting is straightforward.
Let $s_0=0< s_1< ...< s_L <s_{L+1}=\infty$ be free parameters to be optimized and parameterize the powers as
$P_\ell = (e^R-1)/s_\ell$, $\ell=1,...,L$. User $u$ can transmit with power $P_\ell$ if $B=\ell+K(u-1)$,
where $u$ and $\ell$ are chosen to satisfy
\[
u=\arg\max_{k=1,...,K}\{|h_k|^2\} \ {\rm and} \ s_\ell< |h_u|^2\leq s_{\ell+1},
\]
and zero otherwise. 
With this choice of feedback and power levels, successful decoding is always possible
except when $B=0$, thus the throughput is
\begin{align}
    &\eta_{M=1, F=1+KL, K}^{\rm(multilevel cdTDMA-on/off)} \nonumber
  \\&\stackrel{\star}{=}
    \max_{\{s_\ell\}}\log \left( 1+\frac{\overline{P}}{\sum_{\ell=1}^{L}\frac{\mathbb{P}[F=\ell]}{s_\ell}}
    \right)(1-(1-e^{-s_1})^K),
\label{eq:M=1, F=1+KL}
\end{align}
with
\[
\mathbb{P}[F=\ell] = \Pr[s_\ell< \max_{u=1,...,K}\{|h_u|^2\}\leq s_{\ell+1}].
\]
When $L\to\infty$, the multilevel cdTDMA-on/off scheme reduces to ``truncated channel inversion''~\cite{caire_taricco_biglieri:optimal_powercontrol}, which is the optimal in the sense of minimizing the outage probability for a long-term power constraint.

\section{Case $M\geq2$ ($M-1$ possible retransmissions)}
\label{sect:M=2}

In the previous section (see~\eqref{eq:M=1, F=1+KL}) we saw that the throughput increases with $F$.
In this section we evaluate the throughput improvement due to an increase in the number
of retransmissions $M$.  We focus on chTDMA-type policies only, as they emulate the power
allocation that achieves the ergodic water-filling capacity. We shall only consider protocols
for symmetric scenarios; extension to general scenarios is straightforward but tedious.


\begin{figure}
\centering
\includegraphics[width=2.0in,keepaspectratio]{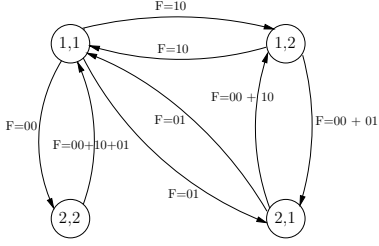}
\caption{Finite-state machine diagram for the cdTDMA+ALO protocol.}
\label{fig:state_machine}
\end{figure}

\begin{figure}
\centering
\includegraphics[width=3.5in,keepaspectratio]{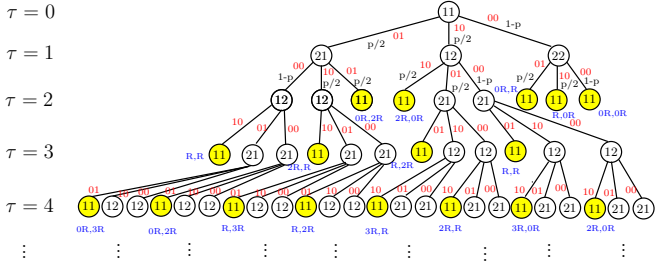}
\caption{Time evolution with time $\tau$ of the cdTDMA+ALO protocol.}
\label{fig:graph}
\end{figure}

\smallskip
\underline{cdTDMA+ALO}:
Here we consider in detail the case $K=M=2$, but the analysis
extends easily.
We start by considering a scheme where a user is scheduled to
transmit when it has the largest fading gain among all users, and its fading is large
enough to grant successful decoding for the available transmit power;
in this case an outage occurs if a user is not scheduled for transmission for $M$ consecutive time slots; the scheme requires $F=1+K$ feedback values.

As shown in~\cite{tuninettiICC2008}, 
by knowing the average reward/decoded rate $\mathbb{E}[\mathcal{R}]$,
the average cost/transmit power $\mathbb{E}[\mathcal{P}]$,
and average inter-renewal time/decoding time $\mathbb{E}[\mathcal{T}]$,
the long-term average throughput is $\mathbb{E}[\mathcal{R}]/\mathbb{E}[\mathcal{T}]$
and the long-term average power is $\mathbb{E}[\mathcal{P}]/\mathbb{E}[\mathcal{T}]$. 
The computation of these quantities if not straightforward in a multiuser channel.
Imagine the protocol as a finite state machine with states:
$(1,1)$: both users are at the first transmission attempt;
$(1,2)$: user 1 is at the first transmission attempt, while user 2 is at the second (and last);
$(2,1)$: user 1 is at the second transmission attempt, while user 1 is at the first;
$(2,2)$: both users are at the second transmission attempt and can not retransmit the same packet any more. A renewal occurs when the system is in the state $(1,1)$.
A state diagram describing the possible transitions among states is given in Fig.\ref{fig:state_machine}, where the possible feedback 
values over two consecutive slots that grant the corresponding transition
are indicated over each arrow.
Given the usual parameterization of rates and powers with the threshold $s>0$,
we can define the probabilities of each feedback value:
$\mathbb{P}[B=0]=\mathbb{P}[\max\{|h_1|^2,|h_2|^2\}\leq s]=(1-e^{-s})^2 = 1-p$, and
by symmetry
$\mathbb{P}[B=1]=\mathbb{P}[B=2] = p/2$.

From the state diagram in Fig.~\ref{fig:state_machine} we can compute the stationary distribution
of the  corresponding aperiodic and irreducible Markov chain, which we denote by $\pi$: $\pi_{i,j}=\mathbb{P}[{\rm state}\ (i,j)]$. With simple algebra, one finds that the average inter-renewal time is 
\begin{equation}
    \mathbb{E}[\mathcal{T}]= \frac{1}{\pi_{1,1}}= 4-p.
\end{equation}
In order to evaluate the average reward, we must understand the evolution of the protocol with time, which is depicted in Fig.~\ref{fig:graph}. In Fig.~\ref{fig:graph} the state $(1,1)$
correspond to renewal events, i.e., the systems starts anew.
Let isolate the right branches starting from the state $(2,2)$ at time $\tau=1$
and compute the accumulated reward. Only the branches labeled with $p/2$ contribute
to the reward as they correspond to successful decoding of rate $R$, that is,
$\mathbb{E}[\mathcal{R}|_{\rm rx}]= R\ p\ (1-p)$.
For the left branches,
the accumulated reward as a function of
the time $\tau$ can be expressed by using the Pascal's triangle as follows:
%
\begin{align*}
     &\sum_{k=0}^{\tau-2}2R {\tau-2 \choose k}(k+2)\left( \frac{p}{2} \right)^{k+2}(1-p)^{\tau-2-k}\\
    &=\left( \frac{p}{2}+(1-p)       \right)^{\tau-2}
    +(\tau-2)\left( \frac{p}{2}+1-p  \right)^{\tau-3}\left( \frac{p}{2}  \right).
\end{align*}

The average reward, not considering the left branches is thus
\begin{align*}
&\mathbb{E}[\mathcal{R}|_{\rm lx}]=2R\left( \frac{p}{2}   \right)^2[\sum_{\tau=2}^\infty 2\left(1-\frac{p}{2} \right)^{\tau-2} +\\
&+  \sum_{\tau=2}^\infty(\tau-2)\left(1- \frac{p}{2}  \right)^{\tau-3}\left( \frac{p}{2}  \right)   ]
= 3\ p\ R.
\end{align*}
By summing the two average reward terms,  we get
\begin{equation}\label{reward}
    \mathbb{E}[\mathcal{R}]=
\mathbb{E}[\mathcal{R}|_{\rm lx}]+\mathbb{E}[\mathcal{R}|_{\rm rx}]=
    Rp(4-p).
\end{equation}
and thus
\begin{align}
&\eta_{M=2,F=3,K=2}^{\rm(cdTDMA+ALO)}
= \max \frac{\mathbb{E}[\mathcal{R}]}{\mathbb{E}[\mathcal{T}]}
=\max_{R>0} Rp.
\label{thr_M=2 alo}
\end{align}
We recognize that the throughput in~\eqref{thr_M=2 alo},
with $R=\log(1+2s/p)$, is the same as the case $M=1$ in~\eqref{eq:TDMA_thr};
therefore a retransmission in this case does not improve performance.
Intuitively, this is so because this protocol is equivalent to the Aloha (ALO) scheme in~\cite{caire_tuninetti:arq_it} (because only one transmission is taken into account for decoding).

\smallskip
\underline{cdTDMA+INR}:
Here we propose to extend the previous cdTDMA+ALO scheme so as to include $L>1$ non-zero transmit power levels; in this case each user adopts the single-user HARQ protocol with INcremental Redundancy (INR)
of~\cite{tuninettiICC2008} when scheduled to transmit; the scheme requires $F=1+LK$ feedback values.  
Let $\eta_{M,F,K=1}^{\rm(INR)}(\overline{P};F_{|H|^2})$ be the throughput of the single-user INR protocol in~\cite[pp.1299]{tuninettiICC2008} with power constraint $\overline{P}$ over a fading channel with fading power gain distributed as $F_{|H|^2}$. The throughput of the cdTDMA+INR protocol is
\begin{align}
&\eta_{M,F=1+KL,K}^{\rm(cdTDMA+INR)}
=\eta_{M,F=L,K=1}^{\rm(INR)}(K\overline{P};F_{\max\{|H_1|^2,...,|H_K|^2\}}),
\label{thr_M inr}
\end{align}
because, in a symmetric network, a user is active for a fraction $1/K$ of the time
when is fading gain is distributed as $F_{\max\{|H_1|^2,...,|H_K|^2\}}$.

\section{Numerical Results}
\label{sect:num}
\begin{figure}
\centering
\hspace*{-1cm}
\includegraphics[width=4.2in,keepaspectratio]{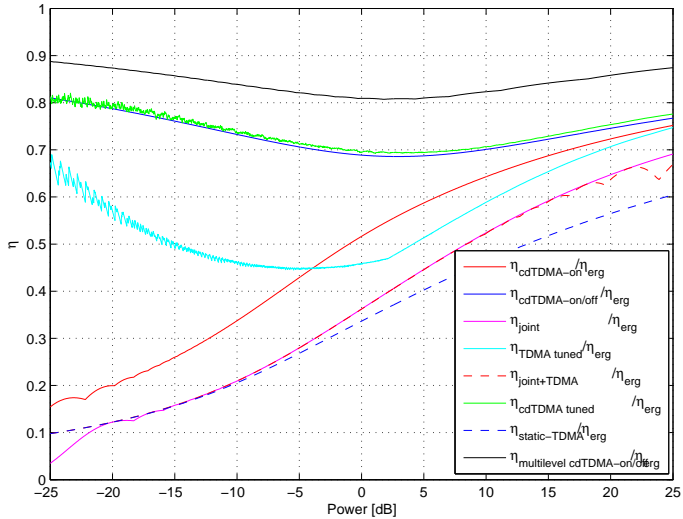}
\vspace*{-1cm}
\caption{Normalized throughput as a function of the average transmit power.}
\label{fig:TDMA}
\end{figure}

In this section we numerically evaluate the throughput
of the different protocols introduced in this paper
for a symmetric iid Rayleigh fading two-user MAC.
Although we did not attempt to optimize the throughput for
a finite $(M,F)$-pair, we now show that small values of $F$
and/or $M$ get close to the water-filling ergodic capacity
(case $(M,F)=(+\infty,+\infty)$).
Fig.~\ref{fig:TDMA} shows the normalized achievable throughput
of the various protocols; the normalization is with respect to
the ergodic water-filling capacity $\eta_{K}^{\rm(ewfc)}$ in~\eqref{eq:erg}.

Case $M=1$ and $F=1$:
From Fig.~\ref{fig:TDMA}, the static TDMA protocol with joint decoding
$\eta_{M=1,F=1,K=2}^{\rm(joint)}$ in~\eqref{eq:joint_nocsi}
has the worst performance at very low SNR.
The static TDMA protocol with no concurrent transmissions
$\eta_{M=1,F=1,K=2}^{\rm (static-TDMA)}$ in~\eqref{eq:tdma_fixed}
is better than $\eta_{M=1,F=1,K=2}^{\rm(joint)}$ at very low SNR because only one user 
is allowed to transmit at any give time, which makes the system interference-free.
At high SNR instead, joint decoding of $\eta_{M=1,F=1,K=2}^{\rm(joint)}$
increases the throughput over $\eta_{M=1,F=1,K=2}^{\rm (static-TDMA)}$.
The protocol $\eta_{M=1,F=1,K=2}^{\rm(joint+TDMA)}$ in~\eqref{eq:joint_single}
combines the advantages  of $\eta_{M=1,F=1,K=2}^{\rm(joint)}$
and  $\eta_{M=1,F=1,K=2}^{\rm (static-TDMA)}$ and outperforms both.
This shows that joint decoding does not substantially improve throughput.

Case $M=1$ and $F\geq2$:
The slight performance enhancement of
$\eta_{M=1,F=2,K=2}^{\rm (cdTDMA-on)}$ in~\eqref{eq:tdma_best}
with respect $\eta_{M=1,F=1,K}^{\rm (static-TDMA)}$ is due to the fact that the best user is always selected for transmission thus reducing the probability of decoding failure.
This shows that scheduling alone (without power control)
is not sufficient to significantly boost performance. 
%
A  performance increase is obtained with 
$\eta_{M=1,F=3,K=2}^{\rm (cdTDMA-on/off)}$ in~\eqref{eq:TDMA_thr} 
with respect to $\eta_{M=1,F=K,K}^{\rm (cdTDMA-on)}$, especially at low SNR.
This is because CSI allows to turn off power when the
channel is in a deep fade; this policy achieves no less than 67\% of the ergodic
wafter-filling capacity with just $1/2\log_2(3)=0.8 < 1$ feedback bits per user.
Similar observations hold for
$\eta_{M=1, F=7, K=2}^{\rm(multilevel cdTDMA-on/off)}$ in~\eqref{eq:M=1, F=1+KL}
this policy achieves no less than 81\% of the
water-filling ergodic capacity with just $1/2\log_2(7)=1.4$ feedback bits per user.
This shows that power control, even based on coarsely quantized CSI, is the
best way to improve performance, especially at low SNR.

Case $M=2$ and $F\geq2$: The highest throughput is obtained with 
$\eta_{M=2,F=7,K=2}^{\rm(cdTDMA+INR)}$ in~\eqref{thr_M inr}, here 
evaluated for $F=7$ so as to compare the benefits of one retransmission
with $\eta_{M=1, F=7, K=2}^{\rm(multilevel cdTDMA-on/off)}$; 
this policy achieves no less than 85\% of the water-filling ergodic capacity
with less that two bits of feedback per user and one retransmission.
Although we only evaluated the throughput for a two-user MAC, we predict
even larger gains from an increase in number of users $K$ because the
protocol $\eta_{M,F=1+KL,K}^{\rm(cdTDMA+INR)}$ is able to combine the
advantages of power control with those of multi-user diversity.

\section{Conclusions}
\label{sect:concl}
We consider the problem of how to best use the limited feedback resources in block-fading MACs:
to provide CSI (to gain multi-user diversity), or to enable HARQ (to gain time diversity)?
We considered the long-term average throughput as a figure of merit and 
we showed that power control (i.e., CSI at the transmitter) seems to be a key factor to improve the throughput in a two-user iid Rayleigh fading MAC with about one bit of feedback per user.
We are currently working on extending the result of this work to systems with larger number of users $K$, larger number of retransmissions $M$, larger number of feedback bits $\log_2(F)$, and different fading statistics to assess the generality of our result.


\bibliographystyle{IEEEtran}
\bibliography{aqricc2011_biblio}

\begin{thebibliography}{10}
\providecommand{\url}[1]{#1}
\csname url@samestyle\endcsname
\providecommand{\newblock}{\relax}
\providecommand{\bibinfo}[2]{#2}
\providecommand{\BIBentrySTDinterwordspacing}{\spaceskip=0pt\relax}
\providecommand{\BIBentryALTinterwordstretchfactor}{4}
\providecommand{\BIBentryALTinterwordspacing}{\spaceskip=\fontdimen2\font plus
\BIBentryALTinterwordstretchfactor\fontdimen3\font minus
  \fontdimen4\font\relax}
\providecommand{\BIBforeignlanguage}[2]{{%
\expandafter\ifx\csname l@#1\endcsname\relax
\typeout{** WARNING: IEEEtran.bst: No hyphenation pattern has been}%
\typeout{** loaded for the language `#1'. Using the pattern for}%
\typeout{** the default language instead.}%
\else
\language=\csname l@#1\endcsname
\fi
#2}}
\providecommand{\BIBdecl}{\relax}
\BIBdecl

\bibitem{book:goldsmith:wireless}
A.~Goldsmith, \emph{Wireless Communications}.\hskip 1em plus 0.5em minus
  0.4em\relax Cambridge University Press, 2005.

\bibitem{knopp_humblet:multiuser_waterfilling}
R.Knopp and P.A.Humblet, ``Information capacity and power control in
  single-cell multiuser communications,'' in \emph{IEEE International
  Conference on Communications, 1995 (ICC '95), 'Gateway to Globalization'},
  vol.~1, Seattle, July 1995, pp. 331--335.

\bibitem{tuninettiICC2008}
J.~Perret and D.~Tuninetti, ``Repetition protocols for block fading channels
  that combine transmission requests and state information,'' in
  \emph{Proceedings of ICC Int. Conf. Comm., ICC2008}, Beijing, China, May
  2008.

\bibitem{Bhashyam}
S.~Bhashyam, A.~Sabharwal, and B.~Aazhang, ``Feedback gain in multiple antenna
  systems,'' \emph{IEEE Trans. on Commun.}, vol.~50, no.~5, pp. 785 -- 798, May
  2002.

\bibitem{tse_hanly_p1}
D.~Tse and S.~Hanly, ``Multiaccess fading channels-part {I}: Polymatroid
  structure, optimal resource allocation and throughput capacities,''
  \emph{IEEE Trans.\ Inform.\ Theory}, vol.~44, no.~7, pp. 2796--2815, November
  1998.

\bibitem{caire_taricco_biglieri:optimal_powercontrol}
G.~Caire, G.~Taricco, and E.~Biglieri, ``Optimum power control over fading
  channel,'' \emph{IEEE Trans.\ Inform.\ Theory}, vol.~45, no.~5, pp.
  1468--1489, July 1999.

\bibitem{tuninettijouiranlACKvsCSI}
D.~Tuninetti, ``On the benefits of partial channel state information for
  repetition protocols in block fading channels,'' in \emph{Submitted to IEEE
  Transactions on Info.Theory}, Jan 2008.

\bibitem{summa_fading_bc}
E.Biglieri, J.Proakis, and S.Shamai, ``Fading channels: information-theoretic
  and communications aspects,'' \emph{IEEE Trans.\ Inform.\ Theory}, vol.~44,
  no.~6, pp. 2619 --2692, Oct. 1998.

\bibitem{caire_tuninetti:arq_it}
G.~Caire and D.~Tuninetti, ``The throughput of {H}ybrid-{ARQ} protocols for the
  {G}aussian collision channel,'' \emph{IEEE Trans.\ Inform.\ Theory}, vol.~47,
  no.~5, pp. 1971--1988, July 2001.

\bibitem{Kim-Skolunt}
T.~T. Kim and M.~Skoglund, ``On the expected rate of slowly fading channels
  with quantized side information,'' \emph{IEEE Trans. on Commun.}, vol.~55,
  no.~4, pp. 820 -- 829, April 2007.

\bibitem{steinber-samai}
A.~Steiner and S.~S. (Shitz), ``Broadcasting with partial transmit channel
  state information,'' \emph{Joint NEWCOM-ARoC Workshop}, Sep. 2006.

\bibitem{davidemsthesis}
D.~Barbieri, ``Transmitter channel state information and repetition protocols
  in multiple access block-fading channel,'' Master's thesis, University of
  Illinois at Chicago, Chicago, IL USA, 2010.

\end{thebibliography}

\end{document}